\documentclass[%
 aip,
 jmp,%
 amsmath,amssymb,
 reprint,%
]{revtex4-1}

\usepackage{graphicx}
\usepackage{epstopdf}
\usepackage{xcolor}
\usepackage{gensymb}
\usepackage{dcolumn}
\usepackage{bm}

\begin{document}

\preprint{AIP/123-QED}

\title[Microwave-regime demonstration of plasmonic non-reciprocity in a flowing two-dimensional electron gas]
{Microwave-regime demonstration of plasmonic non-reciprocity in a flowing two-dimensional electron gas}
\author{Jingyee Chee}
\altaffiliation{now at KLA Corporation}
\affiliation{John A. Paulson School of Engineering and Applied Sciences, Harvard University}

\author{Han Sae Jung}
\affiliation{John A. Paulson School of Engineering and Applied Sciences, Harvard University}

\author{Shannon Harvey}
\altaffiliation{now at Department of Applied Physics, Stanford University}
\affiliation{Department of Physics, Harvard University}

\author{Kenneth West}
\author{Loren Pfeiffer}
\affiliation{Department of Electrical and Computer Engineering, Princeton University}

\author{Amir Yacoby}
\affiliation{John A. Paulson School of Engineering and Applied Sciences, Harvard University}
\affiliation{Department of Physics, Harvard University}

\author{Donhee Ham}
\thanks{Author to whom correspondence should be addressed: \\donhee@seas.harvard.edu.}
\affiliation{John A. Paulson School of Engineering and Applied Sciences, Harvard University}

\date{9 February 2025}

\begin{abstract}
The speed of a plasmonic wave in the presence of electron drift in a conductor depends on the wave's propagation direction, with the wave traveling along the drift (`forward wave') faster than the wave traveling against the drift (`backward wave'). Phenomena related to this plasmonic non-reciprocity---which is relatively more pronounced in two-dimensional conductors than in bulk conductors and could lead to solid-state device applications---have been studied in THz and optical spectral regimes. Here we demonstrate the plasmonic non-reciprocity at microwave frequencies (10 $\sim$ 50 GHz). Concretely, we conduct, at 4K, a microwave network analysis on a gated GaAs two-dimensional electron gas with electron drift ({\it i.e.}, DC current), directly measuring out forward and backward wave speeds via their propagation phase delays. We resolve, for example, forward and backward wave speeds of $4.26 \times 10^{-3} \pm 8.97 \times 10^{-6}$ (normalized to the speed of light). Sufficient consistency between the electron drift speed obtained from the microwave measurement and that alternatively estimated by a DC transport theory further confirms the non-reciprocity. We conclude this paper with a discussion on how to enhance the non-reciprocity for real-world applications, where degeneracy pressure would play an important role. 
\end{abstract}

\maketitle

\noindent When a plasmonic wave (electron density wave) and a DC current (electron drift) are set up together in a conductor, the wave speed depends on its propagation direction \cite{Landau}. That is, if electrons drift at a speed $v_0$ due to a DC current $I_0$, the speed $v_{\text{p},+}$ of a plasmonic wave traveling along the drift (`forward wave') and the speed $v_{\text{p},-}$ of a plasmonic wave traveling against the drift (`backward wave') are given by  
\begin{equation}
v_{\text{p},\pm} = v_{\text{p},0} \pm v_0,
\end{equation}
where $v_{\text{p},0}$ is the plasmonic wave speed in the absence of electron drift. The forward wave is faster, and thus has a larger wavelength at the same frequency. This direction-dependent wave speed, or wave non-reciprocity, can be generally derived by transforming the wave equation from the reference frame moving with the drift to the laboratory frame. It thus is not unique to plasmonic waves in conductors, and can occur, for example, in sound waves in fluid\cite{Landau}. The plasmonic non-reciprocity in conductors, however, may find potential solid-state device applications, such as gain devices where energy is exchanged among the forward wave, backward wave, and current\cite{Landau, Shur}. 

In fact, for any practical ranges of $v_{\text{p},0}$ and $v_0$ in bulk conductors, $v_{\text{p},0} \gg v_0$ holds with $v_{\text{p},0}$ comparable to the speed of light $c$, and thus the plasmonic non-reciprocity goes virtually unnoticed. By contrast, it may fall within the reach of observation in two-dimensional (2D) conductors. For $v_{\text{p},0} $ in 2D conductors can be made far smaller than $c$ due to the inherent nature of the collective electron dynamics in two dimensions\cite{Ham Nature, Ham Nature Nano, Ham Nano Letter, Ham Royal, Ham APL} ({\it e.g.}, $v_{\text{p},0} \approx c/660$ was reported\cite{Ham Nano Letter}), and $v_0$ can be made substantially high in certain 2D conductors endowed with high electron mobility, such as semiconductor 2D electron gas (2DEG) and graphene. Hence in 2D conductors, while $v_{\text{p},0} \gg v_0$ may still typically hold, the gap between the two speeds can be appreciably smaller as compared to bulk conductors. Indeed, the plasmonic non-reciprocity was optically observed by Raman measurement in a 2DEG \cite{Raman} and the possibility of obtaining reflection gain from such plasmonic non-reciprocity in a 2DEG---which may ultimately enable a new class of self-sustained oscillators---was discussed by Dyakonov and Shur\cite{Shur}, with follow-on works reporting THz emissions that may be associated with the gain \cite{SI1,SI2,SI3,SI4}. 

These prior studies on the plasmonic non-reciprocity and associated phenomena were done in the THz and optical spectral regime. Here we conduct a microwave-regime experiment to directly identify forward and backward plasmonic waves by explicitly measuring out their speeds. Concretely, we use a network analysis---2-port scattering parameter ($s$-parameter) measurement---at frequencies of 10 $\sim$ 50 GHz, which can measure a wave speed via the phase delay due to the wave propagation. For the plasmonic wave medium, we use a gated 2DEG strip fabricated from a GaAs/AlGaAs heterostructure grown by molecular beam epitaxy (MBE), where the electron mobility exceeds $10^6$ cm$^2$/V$\cdot$s at 4K.

\begin{figure}[h]
\begin{center}
\includegraphics[width=0.28\textwidth]{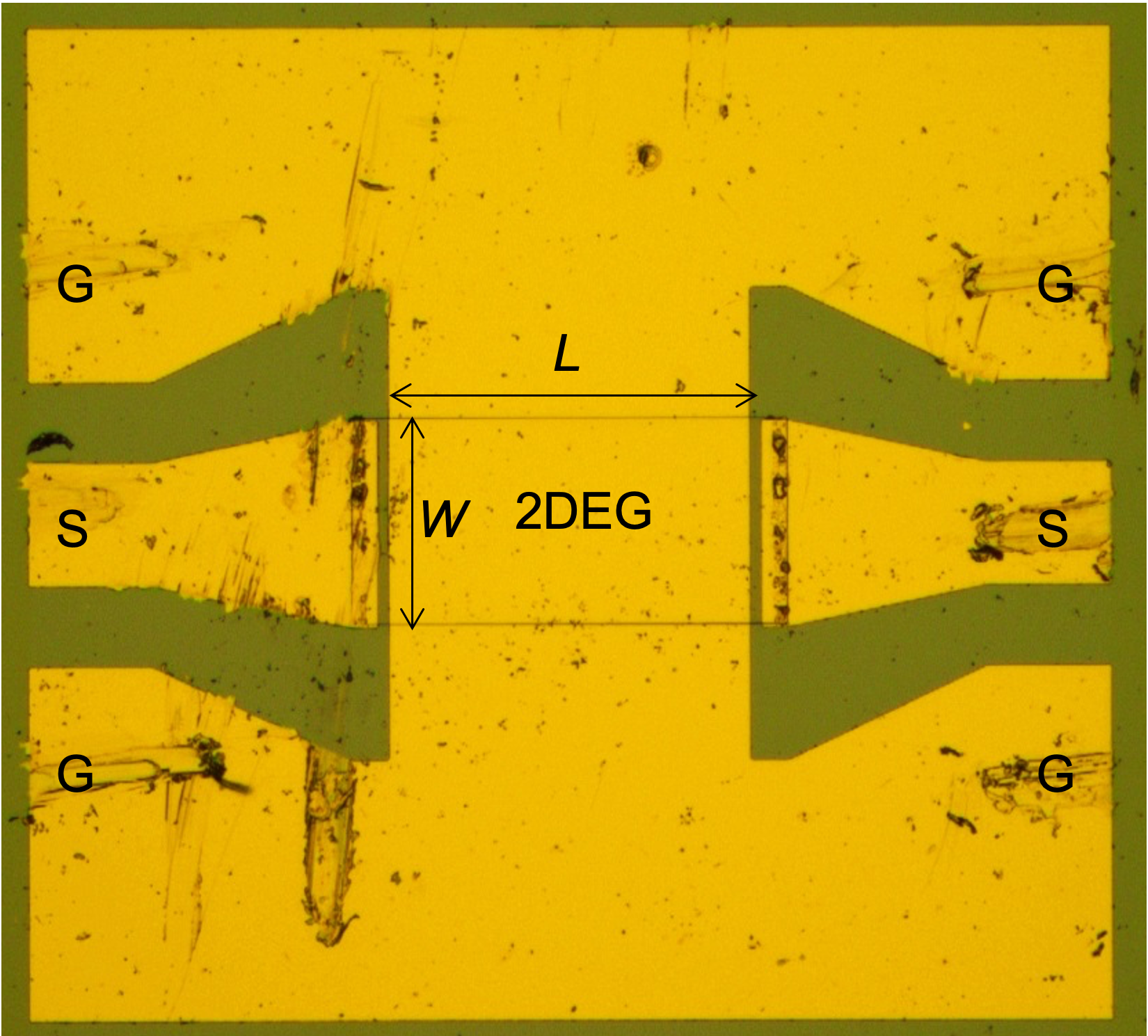}
\caption{\label{fig:setup} A typical device: a gated GaAs 2DEG strip with length $L$ and width $W$, flanked by two planar electromagnetic transmission lines (CPWs). In our actually measured device (not the one in the figure), $L \approx $ 135 $\mu$m and $W \approx 85$ $\mu$m.}
\end{center}
\vspace{-8mm}
\end{figure}

Fig. \ref{fig:setup} shows a typical device, a 2DEG strip flanked at both sides by two planar electromagnetic transmission lines called coplanar waveguides (CPWs). Each CPW comprises a gold signal line and two gold ground lines. The two CPW signal lines are connected through the 2DEG. The ground lines of the two CPWs are directly joined, also extending over to the 2DEG for capacitive coupling (in between the 2DEG and gold ground lie a 75-nm Al$_{0.3}$Ga$_{0.7}$As and a 5-nm GaAs). This capacitively-coupled metal gate serves the microwave ground for the 2DEG. In our actually measured device, $W \approx 85$ $\mu$m (2DEG width) and $L \approx $ 135 $\mu$m (gated 2DEG length; total 2DEG length, including ungated regions, is $\sim$145 $\mu$m). The gated 2DEG is a plasmonic transmission line\cite{Ham Royal}, where  
\begin{equation}
v_{\text{p},0} = (L_\text{k} C)^{-1/2}. \label{bvelocity}
\end{equation}
$L_\text{k}$, the per-unit-length kinetic inductance of the 2DEG, captures the effect of the electrons' collective inertial acceleration (kinetic energy effect), and $C$, the per-unit-length geometric capacitance due to the gating, captures the effect of the Coulomb restoring force (potential energy effect); as the plasmonic wave is an interplay between the Coulomb restoring force that arises from electron density perturbation and the collective electron accelerations due to the restoring force, $L_\text{k}$ and $C$ are key parameters to describe the plasmonic wave\cite{Ham Royal}. Since $L_\text{k}$ in the 2DEG\cite{Ham Royal, Ham Nature} is given by $L_\text{k} = {m^* /(n_0 e^2 W)}$ where $n_0$ is the conduction electron density and $m^*$ is the effective electron mass, Eq. (\ref{bvelocity}) can be re-written as  
\begin{equation}
v_{\text{p},0} = \left({n_0 e^2 \over m^* C_\square}\right)^{1/2}, \label{bvelocitydetail}
\end{equation}
with $C_\square \equiv C/W$ $\approx 0.13$ $\mu$F/cm$^2$ being the per-unit-area geometric capacitance due to the gating. It is this $v_{\text{p},0}$ that is substantially lower than $c$, because $L_\text{k}$ of the 2DEG is large due to the low dimensionality\cite{Ham Royal, Ham Nature}.  

Overall, from left to right of Fig. \ref{fig:setup} lie an electromagnetic transmission line (CPW), a plasmonic transmission line (gated 2DEG), and another electromagnetic transmission line (CPW). A GHz electromagnetic wave launched onto a CPW will excite a GHz plasmonic wave in the gated 2DEG, which will in turn excite an electromagnetic wave onto the other CPW. We perform a network analysis on this setup to measure $s$-parameters, in particular, $s_{12}$ and $s_{21}$, through which we measure propagation phase delays and thus plasmonic wave speeds. 

To fabricate the device, we wet-etch a GaAs/AlGaAs heterostructure wafer to obtain a rectangular mesa that defines the 2DEG boundary. We then deposit, and anneal at 460 $\degree$C, metal alloys (6-nm Ni / 30-nm Au / 60-nm Ge / 20-nm Ni / 150-nm Au) at the two ends of the 2DEG to form Ohmic contacts. Finally, we define the two gold CPWs via photolithography with their signal lines terminated at the two Ohmic contacts. 

The microwave network analysis is performed at 4K on a cryogenic probe station (Lakeshore TTP4). Two ports of a network analyzer (Keysight E8364A) are connected to the two CPWs via two coaxial cables and two GSG probes that land on the far sides of the CPWs. The effects of the cables and probes in the network analysis are calibrated out\cite{marks}. Bias tees internal to the network analyzer are used to apply two DC bias voltages $V_1$ and $V_2$ to the two CPW signal lines. Since we use the gate metal above the 2DEG as both microwave and DC ground, for the gate biasing to tune $n_0$, we use $V_1 = V_2 \equiv V_{\text{bias}}$ for which $I_0 =0$. On the other hand, if we slightly increase $V_1$ and decrease $V_2$ from $V_{\text{bias}}$, we can set up a DC current $I_0$ with a corresponding electron drift speed $v_0 = I_0/ (W e n_0)$, while keeping $n_0$ approximately at the value determined at $V_{\text{bias}}$.  

\begin{figure}[h]
\begin{center}
\includegraphics[width=0.50\textwidth]{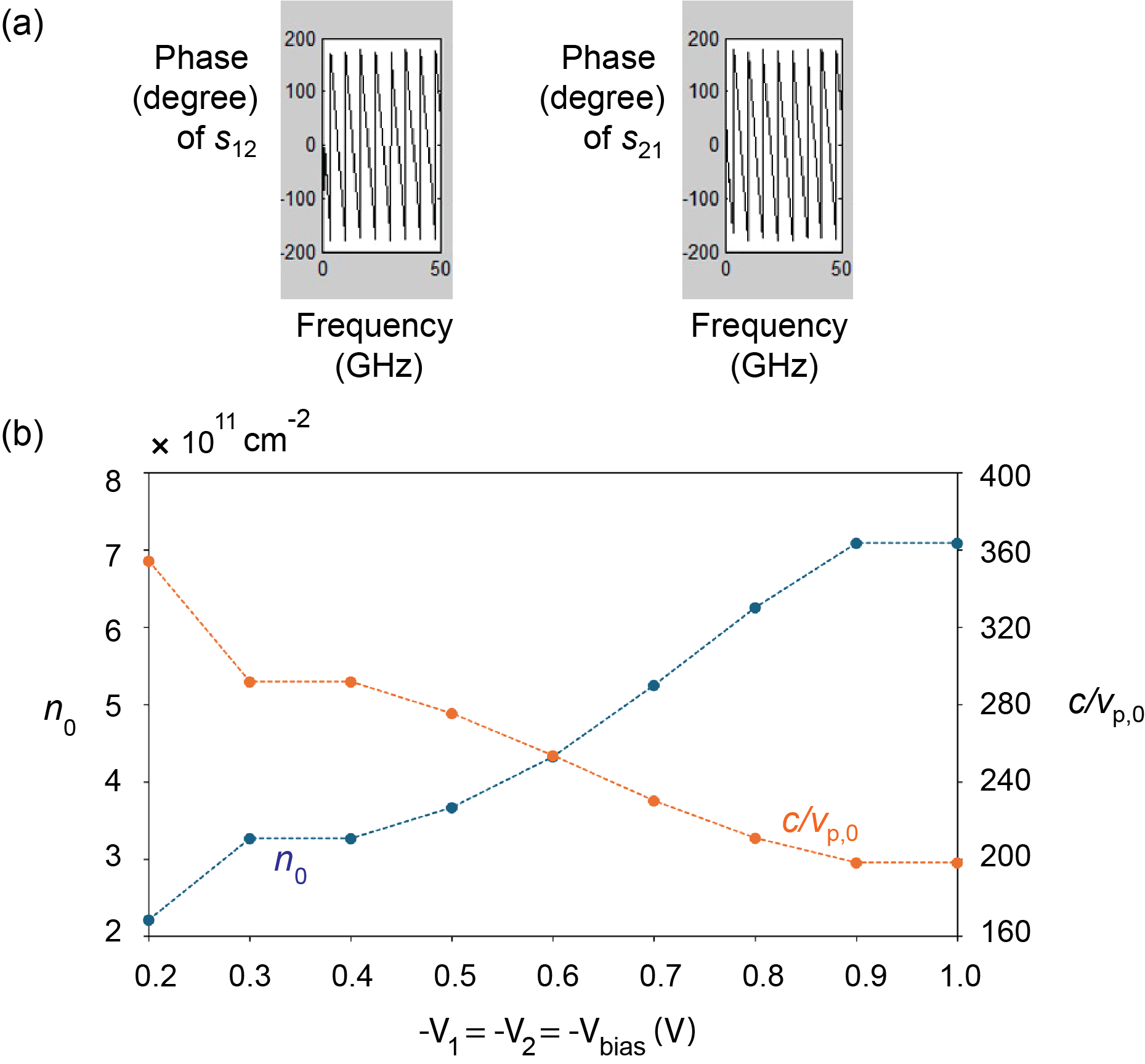}
\vspace{-3mm}
\caption{\label{fig:s} Measurement of reciprocal plasmonic waves ($I_0 = 0$). (a) Phase of $s_{12}$ or $s_{21}$ vs. frequency (10 $\sim$ 50 GHz) for $V_{\text{bias}} \approx -0.2$ V. (b) $c/v_{\text{p},0}$ vs. $|V_{\text{bias}}|$ (orange) and $n_0$ vs. $|V_{\text{bias}}|$ (blue) obtained from the reciprocal plasmonic wave measurement.}
\end{center}
\vspace{-6mm}
\end{figure}

\begin{figure}[h]
\begin{center}
\includegraphics[width=0.46\textwidth]{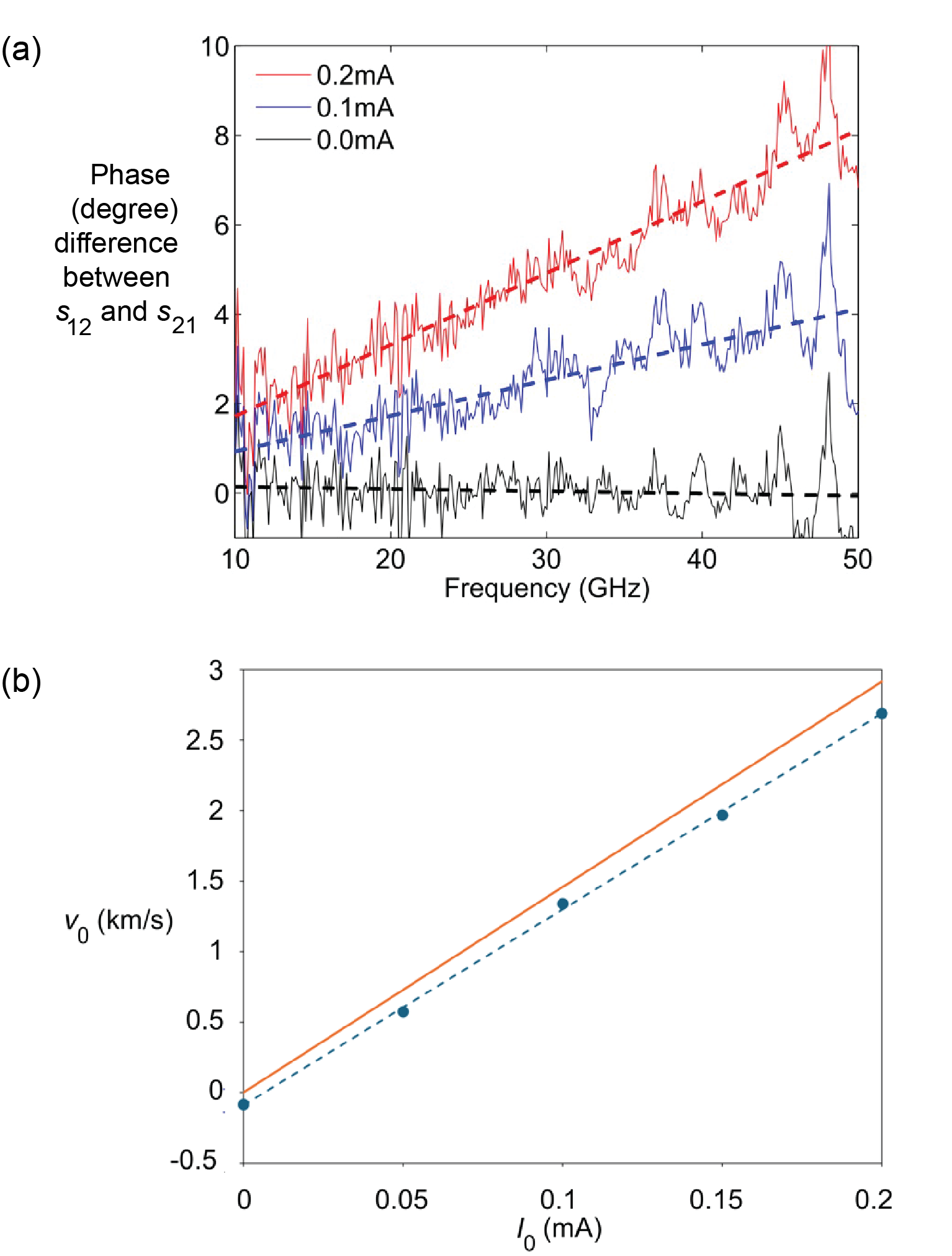}
\caption{\label{highlight} Measurement of non-reciprocal plasmonic waves for various $I_0$ values. (a) Difference between $s_{12}$ and $s_{21}$ phases vs. frequency (10 $\sim$ 50 GHz), shown for select $I_0$ values (the base $V_{\text{bias}}$ from which $V_1$ and $V_2$ are varied to cause $I_0$ is -0.68 V). (b) The $v_0$ vs. $I_0$ data obtained from the microwave-regime measurements (of which Fig. \ref{highlight}(a) is a part) are shown as dots, with the dashed line being their linear fit. The solid line represents the DC-transport formula $ v_0 = I_0 / (n_0 W e)$.}
\end{center}
\vspace{-8mm}
\end{figure}

We first set $I=0$ with $V_1 = V_2 = V_{\text{bias}}$. In this case, the plasmonic non-reciprocity disappears and $s_{12} = s_{21}$, ideally. The phase of either $s_{12}$ or $s_{21}$ is the sum of the phase delay due to the electromagnetic wave propagation through the CPWs with the speed on the order of $c$ and the phase delay $\phi$ due to the plasmonic wave propagation through the gated 2DEG with the speed of $v_{\text{p},0}$. However, as the former is much smaller than the latter with $v_{\text{p},0} \ll c$, the phase of either $s_{12}$ or $s_{21}$ reasonably approximates $\phi$, which is   
\begin{equation}
\phi \approx {2\pi L \over v_{\text{p},0}} f. \label{nonr-phase}
\end{equation}
As seen in an example measurement of Fig. \ref{fig:s}(a) with $V_{\text{bias}} \approx -0.2$ V, the phase of $s_{12}$ or $s_{21}$ indeed linearly grows with frequency in accord with Eq. (\ref{nonr-phase}), where the negative slope in Fig. \ref{fig:s}(a) is the convention of network analysis. By equating the magnitude of the measured slope to $2 \pi L / v_{\text{p},0}$ of Eq. (\ref{nonr-phase}), we obtain $v_{\text{p},0} \approx c /354$ in the case of Fig. \ref{fig:s}(a). The orange dots of Fig. \ref{fig:s}(b) show $v_{\text{p},0}$ so measured at various $V_{\text{bias}}$ (to be exact, the data shown are the slowing factor, $c/v_{\text{p},0}$, of the plasmonic wave speed in comparison to $c$). Figure \ref{fig:s}(b) also shows $n_0$ vs. $V_{\text{bias}}$ (blue dots), where each $n_0$ at a given bias is obtained from the measured $v_{\text{p},0}$ at that bias using Eq. (\ref{bvelocitydetail}). Overall, $n_0$ increases with $|V_{\text{bias}}|$ with an overall slope not too far from $C_\square / e \approx 8.13 \times 10^{11} $ V$^{-1} \cdot $cm$^{-2}$, and $v_{\text{p},0}$ increases with $|V_{\text{bias}}|$ as well, both as expected. Importantly, these measurements confirm the slow plasmonic wave speed $v_{\text{p},0}$ in the gated 2DEG, and provide $n_0$ and $v_{\text{p},0}$ values to utilize in the analysis of the non-reciprocal plasmonic wave measurements presented now.   

We now set up a DC current $I_0$ by increasing $V_1$ and decreasing $V_2$ slightly from $V_{\text{bias}} = -0.68$ V, at which $n_0 \approx 5.07 \times 10^{11}$ cm$^{-2}$ and $v_{\text{p},0} \approx c/235$ are interpolated from the reciprocal plasmonic wave measurement [Fig. \ref{fig:s}(b)]. We then perform microwave network analysis to measure $s_{12}$ and $s_{21}$. Figure \ref{highlight}(a) shows the phase difference between $s_{12}$ and $s_{21}$ as a function of frequency for a few example values of $I_0$. The phase difference between $s_{12}$ and $s_{21}$ would be ideally the same as the difference between the phase delays $\phi_+$ and $\phi_-$ due to the forward and backward plasmonic wave propagation, which can be quantified as 
\begin{eqnarray}
|\phi_+ - \phi_-|  
= \left| {2\pi f L  \over {v_{\text{p},0}+v_0}} - {2\pi f L  \over v_{\text{p},0}-v_0} \right| \approx {4\pi L v_0\over v_{\text{p},0}^2} \times f,
\label{difference}
\end{eqnarray}
where we have used $v_{\text{p},0}^2 - v_0^2 \approx v_{\text{p},0}^2$, as $v_{\text{p},0} \gg v_0$. As seen in Fig. \ref{highlight}(a), the phase difference between $s_{12}$ and $s_{21}$ indeed increases linearly with frequency for a given $I_0$, and furthermore, the slope grows with $I_0$ and thus $v_0$. These observations agree to Eq. (\ref{difference}). Quantitatively, by measuring the slope of the $|\phi_+ - \phi_-|$ vs. $f$ line in Fig. \ref{highlight}(a) for any given $I_0$ and equating it to $4\pi Lv_0/v_{\text{p},0}^2$ in Eq. (\ref{difference}), we can extract $v_0$ vs. $I_0$ data points (to this end, we use the aforementioned $v_{\text{p},0} \approx c/235$), as shown as dots in Fig. \ref{highlight}(b). On the other hand, we draw a theoretical $v_0$ vs. $I_0$ line [Fig. \ref{highlight}(b), solid line], using the DC transport formula of $v_0 = I_0 / (W e n_0)$, for which we use the aforementioned $n_0 \approx$ 5.07 $\times 10^{11}$ cm$^{-2}$. The agreement between the results from the two independent treatments---the data points obtained from the microwave measurements and the theory line obtained from the DC transport consideration---is sufficient enough to confirm the plasmonic non-reciprocity: in Fig. \ref{highlight}(b), between the theory and the data, the $v_0$ values are 21\% off for $I_0 = 0.05$ mA, and less than 10\% off for $I_0 =$ 0.1 mA, 0.15 mA, and 0.2 mA; in the same figure, the slopes of the theory line (solid line) and the linear fit for the data points (dashed line) are less than 5\% off.  One of the sources for the discrepancy between the two results is likely to be the $n_0$ value of $\sim$ 5.07 $\times 10^{11}$ cm$^{-2}$ or the corresponding $v_{\text{p},0}$ value of $\sim c/235$---used in obtaining both results---, which, measured from the reciprocal plasmonic wave measurement more exposed to common-mode noise, is not likely to be the most accurate number. 

Our data analysis has ignored the phase delay through the small ungated regions of the 2DEG ({\it e.g.}, Fig. \ref{fig:setup}). This is well justified, as the plasmonic wave speed in the ungated regions is---due to a different plasmonic wave dispersion relation\cite{Ham Royal}---far larger than $v_{\text{p},0}$ [Eq. (\ref{bvelocitydetail})] of the gated region. On the other hand, to consider the fringing effect of the gate, one could use an effective length $L_{\text{eff}}$ in lieu of $L \approx 135$ $\mu$m ($L_{\text{eff}} > L$) in otherwise the same analysis as in the foregoing. Then the $v_0$ data in Fig. \ref{highlight} would be reduced by a factor of $L/L_{\text{eff}}$, increasing the theory-data discrepancy, but not too significantly. Concretely, given the gate-2DEG distance of only 80 nm, $L_{\text{eff}} \approx 136$ $\mu$m may be a conservative estimate, with which the theory-data discrepancy would remain similar to above. Even for a hypothetical extreme fringing with $L_{\text{eff}} \approx 145$ $\mu$m, the $v_0$ data points in Fig. \ref{highlight} would be off from the transport theory by 14 $\sim$ 27 \%, and the slopes of the theory line and the linear data fit would be 11\% off. These still represent a sufficient agreement, yet such strong fringing is not realistic at all.  

In conclusion, while in bulk conductors where $v_{\text{p},0}$ is hopelessly larger than $v_0$, it is virtually impossible to discern forward and backward plasmonic wave speeds, $v_{\text{p},0} \pm v_0$, in the gated 2DEG with much slower plasmonic wave propagation and high electron mobility, we could tell apart forward and backward waves by measuring out their speeds, $v_{\text{p},0} \pm v_0$. Importantly, however, even in our gated 2DEG, $v_{\text{p},0} \gg v_0$ still holds. For example, $v_0 \approx 2.69 \times 10^3$ m/s at $I_0 = 0.2$ mA [Fig. \ref{highlight}(b)] is $\sim$475 times smaller than $v_{\text{p},0} \approx c/235$. It is the power of the microwave network analysis that resolves the fine difference between $v_{\text{p},0}+v_0$ and $v_{\text{p},0}-v_0$ via phase delay measurement. 

In contrast, the reflection gain\cite{Shur}, $G = (v_{\text{p},0}+v_0)/(v_{\text{p},0}-v_0)$, which could be obtained by open-terminating a gated 2DEG strip and reflecting a forward wave into a backward wave at the termination, would be only $\sim 1.004$ in our case with $I_0 = 0.2$ mA, which is practically too small to observe, for it could be readily masked by Ohmic loss in the 2DEG (the reflection gain here is a voltage gain, the ratio of the amplitude of the oscillating voltage of the 2DEG---proportional to the oscillating charge density in the 2DEG---for the backward wave at the termination to that for the forward wave at the termination). To increase the gain for usability, one is to enhance the non-reciprocity by further decreasing $v_{\text{p},0}$ and increasing $v_0$. To achieve the former, one should go beyond relying only on the large 2D $L_\text{k}$, and increase the gate geometric capacitance [Eq. (\ref{bvelocity})] by decreasing the distance $d$ between the gate and the 2DEG. However, $v_{\text{p},0}$ cannot be made indefinitely small, because the quantum capacitance due to degeneracy pressure\cite{Qcap}---which is always in series with the geometric capacitance, and is ignorable in our present device---will eventually manifest, as the geometric capacitance is made sufficiently large. When the quantum capacitance completely dominates with a small enough $d$, the minimum $v_{\text{p},0}$ for a given $V_{bias}$ would be reached:  
\begin{equation}
v_{\text{p},0,\text{min}}  = (L_\text{k} C_\text{q})^{-1/2} = v_\text{F}/\sqrt{2}. \label{minimum}
\end{equation}
Here $C_\text{q} = me^2W/(\pi \hbar^2)$ is the per-unit-length quantum capacitance and $v_\text{F}$ is the Fermi velocity for the 2DEG. Since $v_\text{F}$ is tunable in the 2DEG by adjusting $n_0$ via $V_{\text{bias}}$, if $v_{\text{p},0}$ is made close enough to $v_{\text{p},0,\text{min}}$ by sufficiently reducing $d$, one may be able to attain a practically large enough reflection gain. The reduction of $d$ to the point of making quantum capacitance dominant would be a practical challenge however, and also there may arise complex tradeoffs in association with tuning down $v_\text{F}$.   

The authors thank Army Research Office (W911NF-17-1-0574 to Harvard), Air Force Office of Scientific Research (FA9550-13-1-0211 to Harvard), and Gordon and Betty Moore Foundation (EPiQS Initiative GBMF9615.01 to Loren Pfeiffer, Princeton) for support. Device fabrication was performed in part at the Harvard Center for Nanoscale Systems (CNS).

\vspace{-3mm}
\section*{Author Declarations}
\vspace{-4mm}
\subsection*{Conflict of Interest}
\vspace{-4mm}
\noindent The authors have no conflicts to disclose.
\vspace{-7mm}
\subsection*{Author Contributions}
\vspace{-4mm}
\noindent Jingyee Chee and Han Sae Jung contributed equally to this paper.
\smallbreak
\noindent{\bf Jingyee Chee:} Conceptualization (equal); formal analysis (equal); investigation (lead). 
{\bf Han Sae Jung:} Formal analysis (equal); writing - original draft (lead). 
{\bf Shannon Harvey:} Investigation (supporting). 
{\bf Kenneth West:} Resources (equal). 
{\bf Loren Pfeiffer:} Resources (equal). 
{\bf Amir Yacoby:} Supervision (supporting).  
{\bf Donhee Ham:} Supervision (lead); conceptualization (equal); funding acquisition (lead); formal analysis (equal); writing - review and editing  (lead). 

\vspace{-3mm}
\section*{Data Availability}
\vspace{-4mm}
\noindent The data that support the findings of this study are available within the article.

\vspace{-3mm}
\section*{Credit Lines}
\vspace{-4mm}
\noindent The following article has been submitted to {\it Applied Physics Letters}. Copyright (2025) Jingyee Chee et al. This article is distributed under a Creative Commons Attribution-NonCommercial 4.0 International (CC BY-NC) License.

\end{document}